\documentclass[10pt,a4paper]{article}
\usepackage[utf8]{inputenc}
\usepackage[T1]{fontenc}
\usepackage{amsmath}
\usepackage{amsfonts}
\usepackage{amssymb}
\usepackage{graphicx}
\usepackage{systeme}
\usepackage[margin=1.15in]{geometry}
\graphicspath{{./}}
\author{T. Nassar, S. Ephrem}
\title{Optimal allocation using the Sortino ratio}
\begin{document}
\maketitle{}
\bibliographystyle{unsrt}
\section{Kelly criterion revisited}
\paragraph{} In 1956, while working at AT\&T, Kelly published a seminal paper inconspicuously entitled "A new interpretation of information rate" \cite{doi:10.1002/j.1538-7305.1956.tb03809.x}. The paper's avowed goal was to provide a betting strategy for a gambler who receives information (through some channel) related to her bets. For instance, suppose the gambler is betting on the outcomes of a coin toss. Prior to each such toss, the gambler receives a communication from a secret benefactor that tells her which way she should bet. If this benefactor is an oracle capable of predicting the future exactly, then clearly, by betting all of her money every time as per the instructions of the benefactor, the gambler can achieve great fortune. This becomes considerably more complicated if the benefactor is not an oracle and therefore has a certain probability p of being right. In this case, Kelly asks what would be the optimal proportion $\Theta$ of the gambler's money that should be bet every time given the probability p. 
\paragraph{} To this end, we observe that if the benefactor's communication is correct then, at time t+1 the gambler's money increases by $(1+\Theta)V_{t}$ where $V_{t}$ is the amount of money the gambler had at time t and if the benefactor is wrong, the gambler's money decreases to $(1-\Theta)V_{t}$ where we have assumed the odds are 1:1. Hence, after a certain time period T, the gambler's fortune would be given by:

\begin{equation}\label{key}
V_{t} = (1+\Theta)^{W}(1+\Theta)^{L}V_{0} = 
        (1+\Theta)^{W}(1+\Theta)^{T-W}V_{0}
\end{equation}
where w is the number of wins and L = T - W is the number of losing bets.
\paragraph{} 
From (1) it is easy to see that if $W\cong T$ (i.e. the benefactor is right most of the time), the value $V_{t}$ will continue to grow in an unbounded way the longer the gambler keeps playing the game. Instead of dealing with such divergent quantities which, due to their divergence as $T \rightarrow +\infty$ become impossible to compare or optimise over, Kelly considers the logarithm of (1):
\begin{equation}
G = \lim_{T\rightarrow\infty}  \dfrac{1}{T} \log\left(\dfrac{V_{T}}{V_{0}}\right) = (1-p)\log(1-\Theta) + p\log(1+\Theta)
\end{equation}
where :
\begin{equation}
p = \lim_{T\rightarrow\infty}  \dfrac{W}{T}
\end{equation}
\\
i.e. $p$ is the probability that the benefactor is right
\newpage
\paragraph{} The quantity defined in (2) bears a relationship to channel capacity in information theory but it also has a much simpler financial interpretation: it is the internal rate of return (IRR) on the betting/investment strategy considered. The proportion of the portfolio invested can now be determined in a straightforward way by maximising G over $\Theta$ :
\begin{equation}
\Theta = 2p - 1
\end{equation}
Clearly, from (4) we note that if $ p < 0.5 $, we get a negative $\Theta$ which implies that we should never do any betting in that case. That's perfectly intuitive since having $ p < 0.5 $ means that our benefactor is more wrong than right and as such is not very trustworthy.
\paragraph{}
The Kelly criterion as set out in equation (4) and justified in (1-3), applies to the case where the decisions available to the gambler are binary and where the odds are 1:1. This need not be the case and the criterion can be generalised to account for non-binary choices and odds different from 1:1 as discussed in \cite{doi:10.1002/j.1538-7305.1956.tb03809.x}. More relevant to us in this paper is the risk associated with the use of a strategy based on the Kelly criterion. In equation (2) one can prove, using the central limit theorem, that G converges almost surely to $(1-p)\log(1-\Theta) + p\log(1+\Theta)$ as $T\rightarrow \infty$:\\ 

$P(\hat{G}=G)=\delta (G-(1-p)\log(1-\Theta)-p \log (1+\Theta))$\\

This fact is sometimes erroneously taken to mean that  use of the Kelly criterion entails no risk. In addition, as pointed out in \cite{breiman1961}, Kelly's criterion minimizes the time needed to reach a certain wealth level. However, as pointed out in \cite{10.2307/61075}, the innocuous limit $T\rightarrow \infty$ is an idealisation that does not approximate reality in a satisfactory way. 

\paragraph{}
To show this, let's denote by $\hat{\eta} \in \{-1,1\}$ whether the decision communicated by the benefactor to the gambler at some time t is correct $(\hat{\eta} = 1)$ or incorrect $(\hat{\eta} = -1)$. The gambler's fortune after T steps is then given by:
\begin{equation}
	\hat{V}_{T} = V_{0} \prod_{t=1}^{T}(1 + \hat{\eta}_{t} \; \Theta)
\end{equation}
so that:
\begin{equation}
\begin{split}
	\hat{G}_{T} &= \dfrac{1}{T} \log\left(\dfrac{\hat{V}_{T}}{V_{0}}\right)\\
	 &= 
	\dfrac{1}{T}\sum_{t=1}^{T}\log(1 + \hat{\eta} \; \Theta)\\
	 &= 
	\dfrac{1}{2}\log(1-\Theta^{2}) + \dfrac{1}{2T}\log\left(\dfrac{1+\Theta}{1-\Theta}\right)\sum_{t=1}^{T}\hat{\eta}_{t}
\end{split}
\end{equation}

\paragraph{}
Kelly refers to the quantity $\hat{G}_{T}$ as the rate of growth of the gambler's fortune. It corresponds to what one would call an internal rate of return in terms of trading strategies. The expected value of $\hat{G}_{T}$ is given by equation (2) and is maximised by equation (4). However, unlike the limit $\hat{G} = \lim_{T\rightarrow\infty} \hat{G}_{T}$, maximising $\hat{G}_{T}$ targets returns but does not take risk into account. In particular, we can, for instance, calculate the Sharpe ratio as follows:
\begin{equation}
\begin{split}
	E\left[ \hat{G}_{_{T}}^{^{2}}\right] - E\left[ \hat{G}_{_{T}}\right]^{2} &= \dfrac{1}{4\;T^{2}}\log^{2}\left(\dfrac{1+\Theta}{1-\Theta}\right)
	(E\left[\sum_{t,s=1}^{T} \hat{\eta_{t}}\hat{\eta_{s}}\right] - T^{2}(2p-1)^{2}) \\
	&= \dfrac{1}{4\;T^{2}}\log^{2}\left(\dfrac{1+\Theta}{1-\Theta}\right)
	(T + T(T-1)(2p-1)^{2}-T^{2}(2p-1)^{2}) \\ \\
	&= \dfrac{1}{T}\log^{2}\left(\dfrac{1+\Theta}{1-\Theta}\right)p(1-p)
\end{split}
\end{equation}
which gives for the Sharpe ratio :
\begin{equation}
S(p, \Theta) = \sqrt{\dfrac{T}{p(1-p)}}\left(p- \dfrac{1}{2} + \dfrac{1}{2}\;\dfrac{\log(1-\Theta^{2})}{\log\left({\dfrac{1+\Theta}
		{1-\Theta}}\right)}\right)
\end{equation}
Inserting allocation (4) in Kelly's criterion :
\begin{equation}
	S(p, 2p - 1) = \sqrt{\dfrac{T}{p(1-p)}}\left(p- \dfrac{1}{2} + \dfrac{\log\left(4(1-p)p\right)}{2\log\left({\dfrac{p}
			{1-p}}\right)}\right)
\end{equation}
Removing the "scaling factor" $\sqrt{T}$ from (7), we can easily plot the behaviour of the Kelly Sharpe ratio against the probability p

\begin{figure}[h]
	\includegraphics[width=10cm]{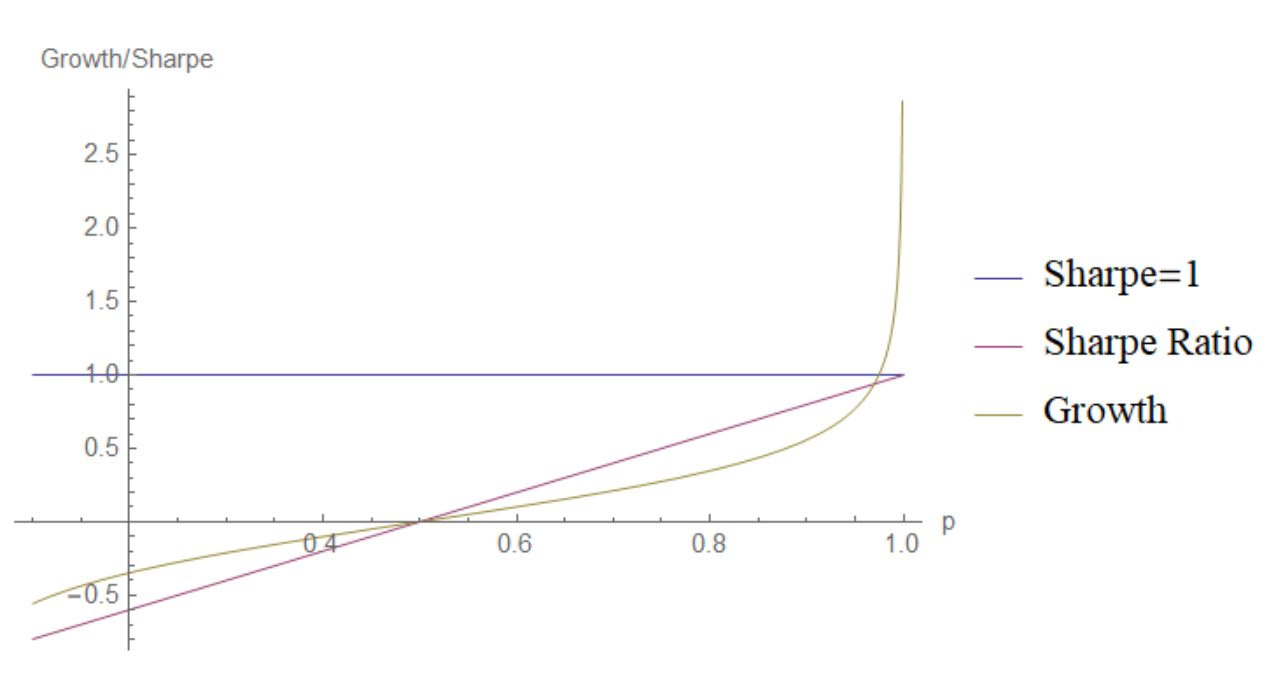}
	\centering
	\caption{Kelly Sharpe Ratio}
\end{figure}

\paragraph{}
Unless the probability p > 0.975, the Sharpe ratio remains below 1 which, in itself, is a fairly low Sharpe ratio. If, instead of using the allocation proposed by Kelly, we try to maximise the Sharpe ratio (as given in (8)) directly, we get:
\begin{equation}
	\partial_{\Theta}S(p, \Theta) = 0 \implies \log(1-\Theta^{2}) + \Theta\log\left({\dfrac{1+\Theta}{1-\Theta}}\right) = 0
\end{equation}
In the following, we will propose a different method where the aim is to try and maximise a different risk-adjusted method, the Sortino ratio, to obtain the optimal allocation/bet. We will then proceed to testing the strategy that consists of optimal amount every day for a certain period of time on the Dow Jones data. Finally, we will provide a summary and a description of future work.

\section{Sortino ratio}
\paragraph{}

The theory of optimal asset allocation is old and well researched. A general and key tenant in optimal asset allocation is to strike a balance between
expected utility maximization and risk minimization. Traditionally, risk is measured by the variance of the returns so that one is trying to optimise:\\
\\
\(\left.\mathcal{L} = E\left[U\left(\hat{r}_T,\Theta \right)\right]-\frac{\lambda }{2}\left(E\left[\hat{r}_T^2\right.\right)-E\left[\hat{r}_T\right]{}^2\right)\)\\
where \(\hat{r}_T\) is the return of a given investment, $\Theta $ is the allocation vector and $\lambda $ parametrises the risk aversion of individual
investors.\\
\\
The first term is the utility and the second is the risk. Different utilities lead to different allocations and in fact Kelly's criterion corresponds basically to the "commonly preferred constant relative risk aversion" as it degenerates to a log-utility function (see \cite{BazPimco} for a comprehensive overview).
The Sharpe ratio: \(S_T=\frac{E\left[\hat{r}_T\right]-\mu }{\sqrt{\left.E\left[\hat{r}_T^2\right.\right)-E\left[\hat{r}_T\right]{}^2}}\) basically corresponds to the case where the utility function is linear and it penalises risk through the use of standard deviation. 
\\\\
The idea is that if this standard deviation goes to zero, the returns on the proposed investment strategy become certain. This suffers from two problems:\\\\
1. Although the return becomes certain as the standard deviation goes to zero, that return may be quite low. Since dividing any number by zero gives infinity, the use of the Sharpe ratio in the case of a very small standard deviation can be very misleading when comparing different strategies.\\\\
2. Standard deviation captures the fluctuations of a random variable around its mean regardless of whether these fluctuations are above or below the mean. In risk terms, the standard deviation penalises both the good returns and the bad! When the distribution of the returns is not symmetrical about the mean (whether that distribution is Gaussian or not doesn't matter), the Sharpe ratio will again be misleading as a performance gauge of different strategies.
\\\\
The Sortino ratio, proposed in 1980 by Frank Sortino \cite{Sortino}, addresses the two problems of the Sharpe ratio (see \cite{SharperRatio}) as follows. Let's define:
\begin{equation}
	\Phi = \dfrac{E\left[\hat{G_{T}}\right]-\mu}
	{\sqrt{E\left[\min^{2}\left(\hat{G_{T}} - \mu,0\right)\right]}}
	= \dfrac{(1-p)\log{(1-\Theta)} + p\log{(1+\Theta)} - \mu}
	{\sqrt{E\left[\min^{2}\left(\hat{G_{T}} - \mu,0\right)\right]}}
\end{equation}
where $\mu$ is a desired rate of return\footnote[1]{The desired return $\mu$ can of course be also set to the "risk free" rate prevalent for the investment horizon being considered, e.g. some Libor or Fed funds rate.} The numerator will be positive if the expected return on the proposed strategy exceeds the desired rate. The novelty is in the denominator. The function $\min\left(\hat{G_{T}} - \mu,0\right)$ picks up the cases where the strategy is achieving returns below the desired rate of return so that the denominator $\mathcal{D} = \sqrt{E\left[\min^{2}\left(\hat{G_{T}} - \mu,0\right)\right]}$ of the Sortino ratio becomes the standard deviation of the "losses" of the strategy. Therefore, the Sortino ratio will penalise returns that are inordinately skewed to the downside.

\paragraph{}
Now we are in a position to calculate the Sortino ratio in the investment/betting setting used by Kelly. Noting that using the function $\min\left(\hat{G_{T}} - \mu,0\right)$ amounts to summing over configurations that satisfy:
\begin{equation}
	\dfrac{1}{2}\log{1-\Theta^{2}} + \dfrac{1}{2T}\log{\left(\dfrac{1+\Theta}{1-\Theta}\right)
	\sum_{t=1}^{T}\hat{\eta_{t}} < \mu} \implies \hat{x_{T}}
	< \min{(T, T_{max})}
\end{equation}
\\
where $\hat{x_{T}} = \sum_{t=1}^{T}\hat{\eta_{t}}$ and $T_{max} = \min{\left(T, \; 
	\dfrac{T(2\mu-\log{(1-\Theta^{2})})}{\log{\left(\dfrac{1+\Theta}{1-\Theta}\right)}}\right)}$ \\ since we also have $\hat{x_{T}} \leq T$ \\\\
\newpage
the denominator becomes :
\begin{equation}
\begin{split}
\mathcal{D}^{2} = 
\sum_{x=0}^{T_{max}}
&\left( \left( \dfrac{1}{2}\log{(1-\Theta^{2})}+\dfrac{1}{2T}\log{\left(\dfrac{1+\Theta}{1-\Theta}x\right)}\right)^{2} \right.\\
&\left. \times\;\dfrac{T!}{x!\;(T-x)!} \; p^{x} (1-p)^{t-x} \right)
\end{split}
\end{equation}
\paragraph{}
Equation (12) also entails that $\forall \; \mu>\log(1+\Theta)$, $\mathcal{D}$ remains the same. Equation (13) on the other hand, horrible as it looks, has a closed form solution, albeit one that uses certain incomplete beta functions : 
\begin{equation}
\begin{split}
	D =
	\dfrac{T!}{a!(T-a-1)!}\left( \right.&p^{a-1}(1-p)^{-a+T-2} \left(-a\alpha+ a \;p(T-1) + b(p-1)p\right)\\
	&+ c B_{1-p}(T-\alpha, \alpha+1)\left.\right)
\end{split}
\end{equation}
where $\alpha$, a, b and c are given in the appendix. To avoid cumbersome calculations, the derivation of (14) is relegated to the appendix. In the following section we shall investigate the behaviour of the allocation based on maximisation of the Sortino ratio relative to the different parameters involved. 

\section{Behaviour with p}
\paragraph{}
As the probability of a {``}win{''} increases, we expect that both the optimal allocation $\Theta $ and the corresponding Sortino ratio increase.
This is indeed the case as figures 1 and 2 (where we have used \(T=90 \text{ and } \mu =0.02=2\%)\) below show:
\begin{figure}[h]
	\includegraphics[width=9cm]{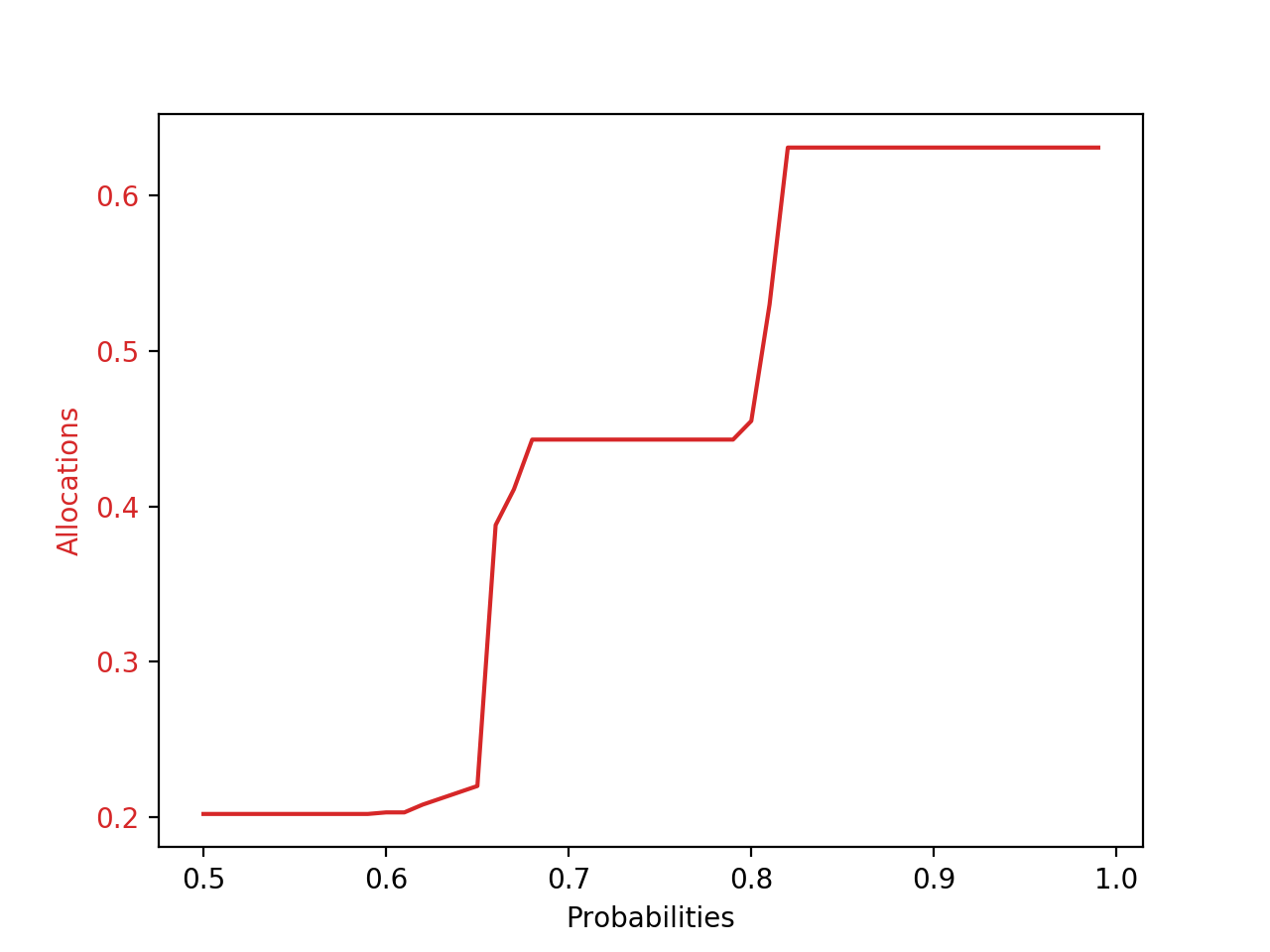}
	\centering
	\caption{Optimal allocations versus probabilities}
\end{figure}
\\
\newpage
The stairlike behaviour is also to be expected due to the denominator of the Sortino ratio which is a discontinuous function. The behaviour of the Sortino ratio itself is quite telling:
\begin{figure}[h]
	\includegraphics[width=9cm]{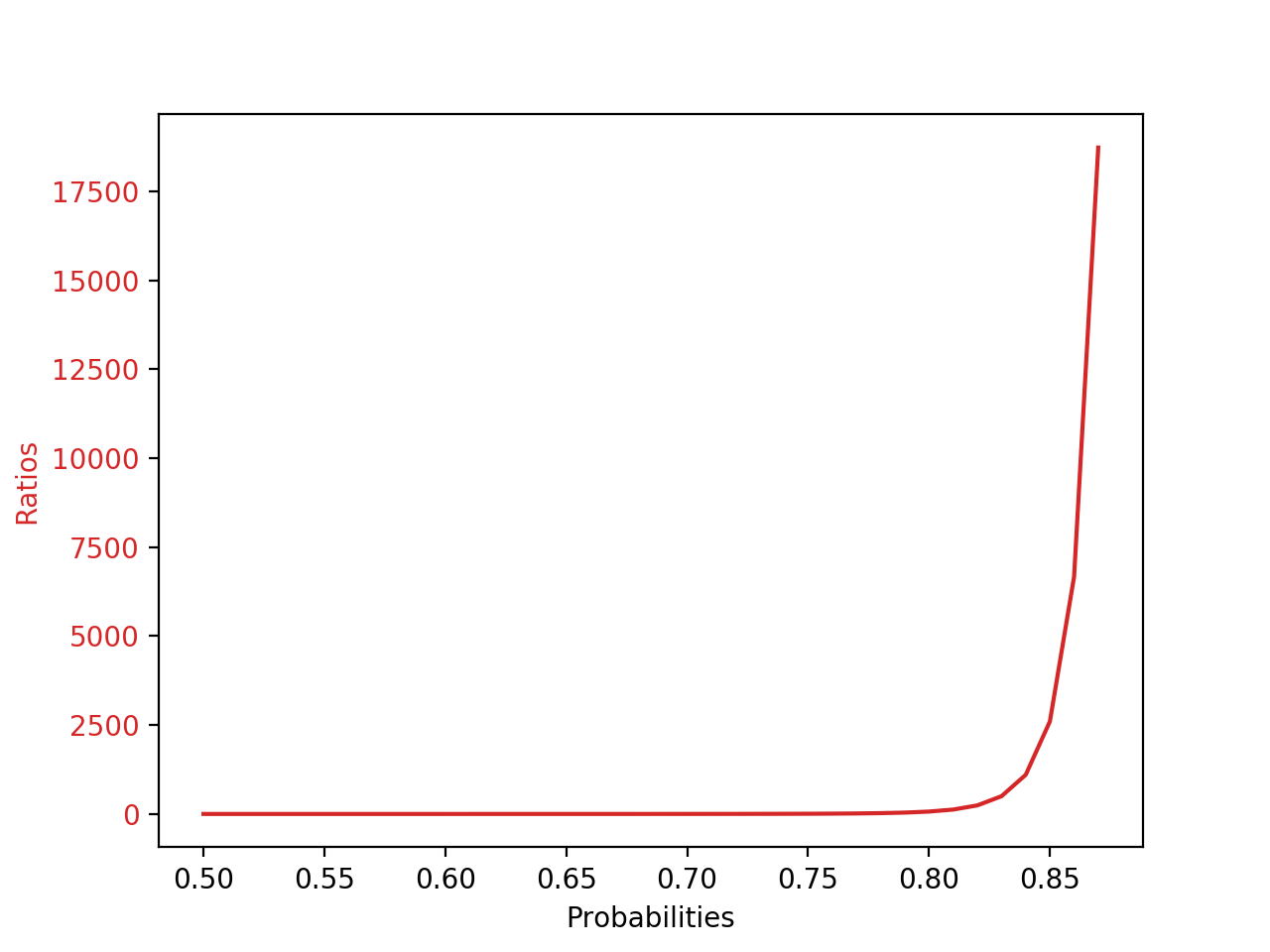}
	\centering
	\caption{Optimal Sortino ratios versus probabilities}
\end{figure}
\\
The previous figure shows that the Sortino ratio doesn't go beyond 2 (which is usually seen as a "good" minimal value) unless p>83.55\% which is very high.
\section{Behaviour with $\mu$}
\paragraph{}
The behaviour with $\mu $ is fairly complicated. Remembering that $\mu $ is a desired level of returns (for instance, a risk free rate) and that
the Sortino ratio punishes return fluctuations that are below the desired level, it seems natural that the Sortino ratio would diminish as $\mu $
increases. This is indeed the case as shown by the graph below (where we have taken \(p=0.72 \text{ and } T = 90)\):
\begin{figure}[h]
	\includegraphics[width=9cm]{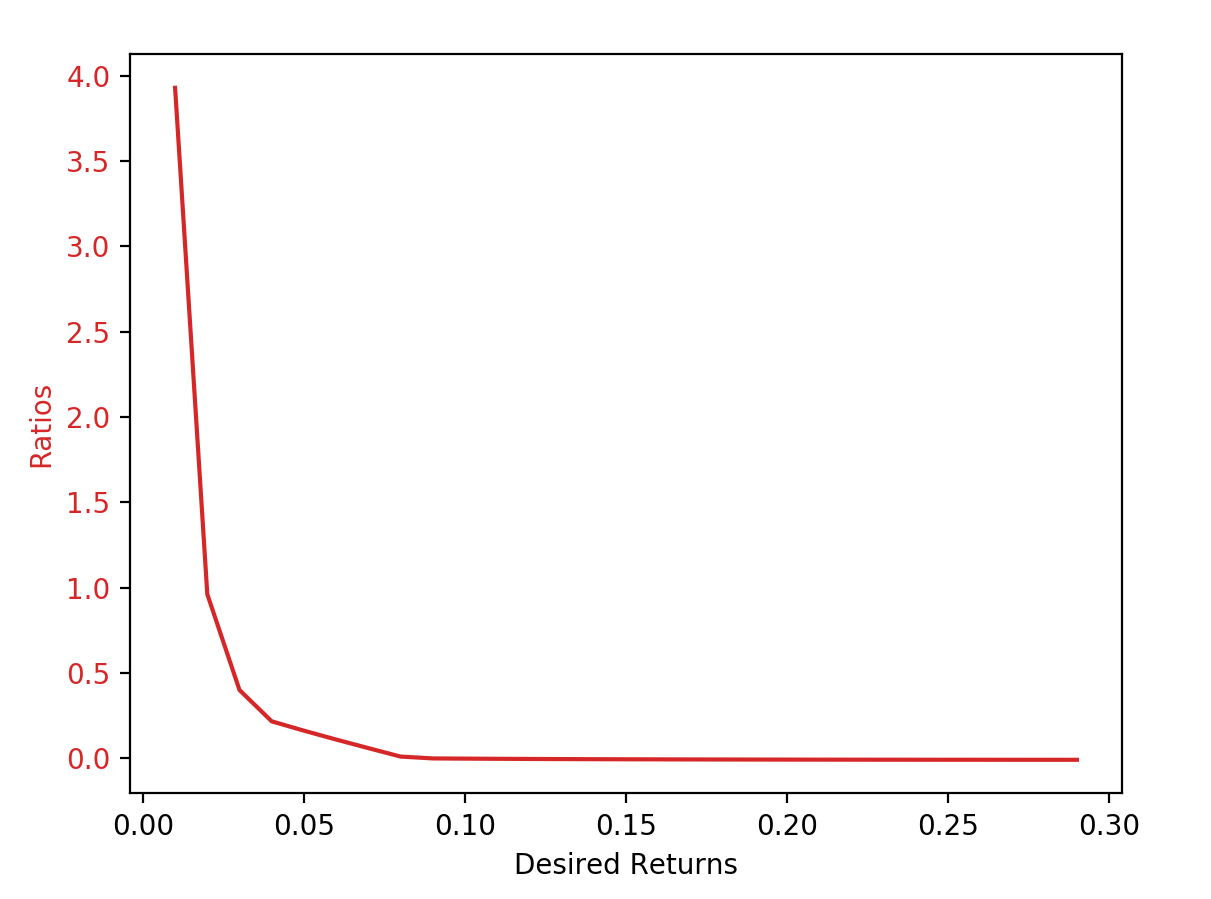}
	\centering
	\caption{Optimal Sortino ratios versus desired returns}
\end{figure}
\newpage
\section{Results}
To showcase how the previous analysis can be applied in real life, we will use the Dow Jones closing prices from 29-01-1985 to 28-08-2019. The period was chosen in such a way that it includes multiple economic and financial shocks and bull markets to illustrate the robustness of the method in different settings. We simulate signals based on the Dow Jones  such that:
\\ \\ $P\left[\hat{s_{t}} = sign(p_{t+1} - p_{t})\right] = p = 0.6$
\\
\\ where $sign(x) = \begin{cases} 1 &\mbox{if } x > 0 \\
-1 & \mbox{if } x < 0 \\ 0 & \mbox{if } x = 0\end{cases}$
\\ 
\\ This can be done by setting :
\\ \\ $\hat{s_{t}} = sign(p_{t+1} - p_{t})\; \hat{\xi_{t}} $
\\ where $\hat{\xi_{t}}\in \{-1, 1\}$ and $\hat{\xi_{t}} \sim p$ since $P\left[\hat{s_{t}} = sign(p_{t+1} - p_{t})\right] = P\left[\hat{\xi_{t}} = 1\right] = p$

\paragraph{}
We repeat each simulation \(L\) times so we denote the generated signals by \(\hat{s}_t^{(a)} \text{where} a=1,\text{..},L\). The strategy consisted
of selling the optimal allocation \(\Theta _*(p,\mu ,T)\) (where \(T=100 \text{ and } \mu =3\%)\) when \(\hat{s}_t^{(a)}=-1\) and buying when \(\hat{s}_t^{(a)}=1\). For each date in the Dow Jones series, a trade was initiated based on the signal $\hat{s_{t}}$ and unwound after T days. The return for each such trade was then calculated and pooled into a series of returns obtained from the strategy. This series was then used to calculate the resulting expected return and the corresponding Sortino ratio. The simulations are then averaged to estimate the probability density of the returns.  In this paper, we have used L = 20,000. The resulting returns for each day are then averaged to estimate the distribution of (annualised) returns. This is shown in figure 4 below: 
\begin{figure}[h]
	\includegraphics[width=12cm]{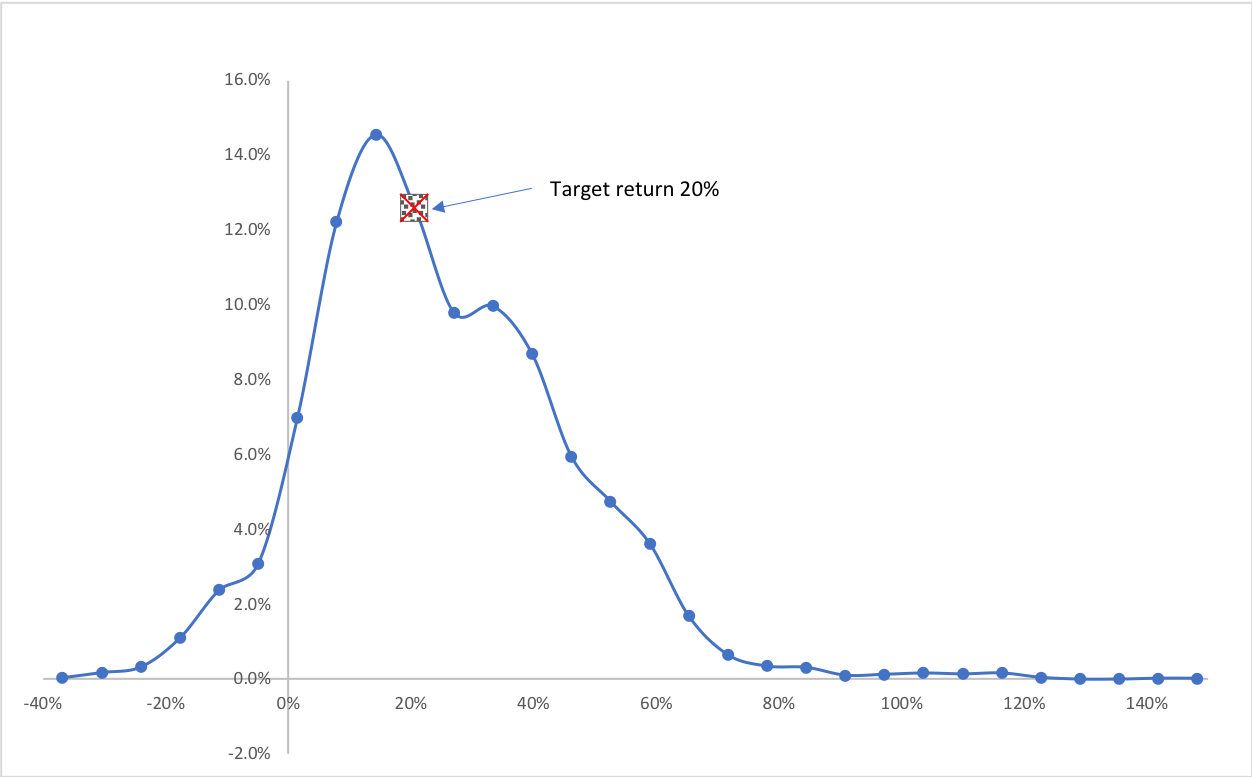}
	\centering
	\caption{Return probabilities from Dow Jones simulations}
\end{figure}
\newpage
The average return is 24.74\% and the Sortino ratio is 2.29. The Sharpe ratio on the other hand is 0.15. This is understandable since, as figure 5 shows, is overwhelmingly skewed towards positive returns and as such, the Sortino ratio will be high whereas the Sharpe ratio will be low. 
Perhaps more interesting than the returns showcased above is the behaviour of this strategy in different market conditions. To test this, we take the series of simulated returns and we plot it against time, highlighting the periods of different financial crisis:\\

\begin{figure}[h]
	\includegraphics[width=12cm]{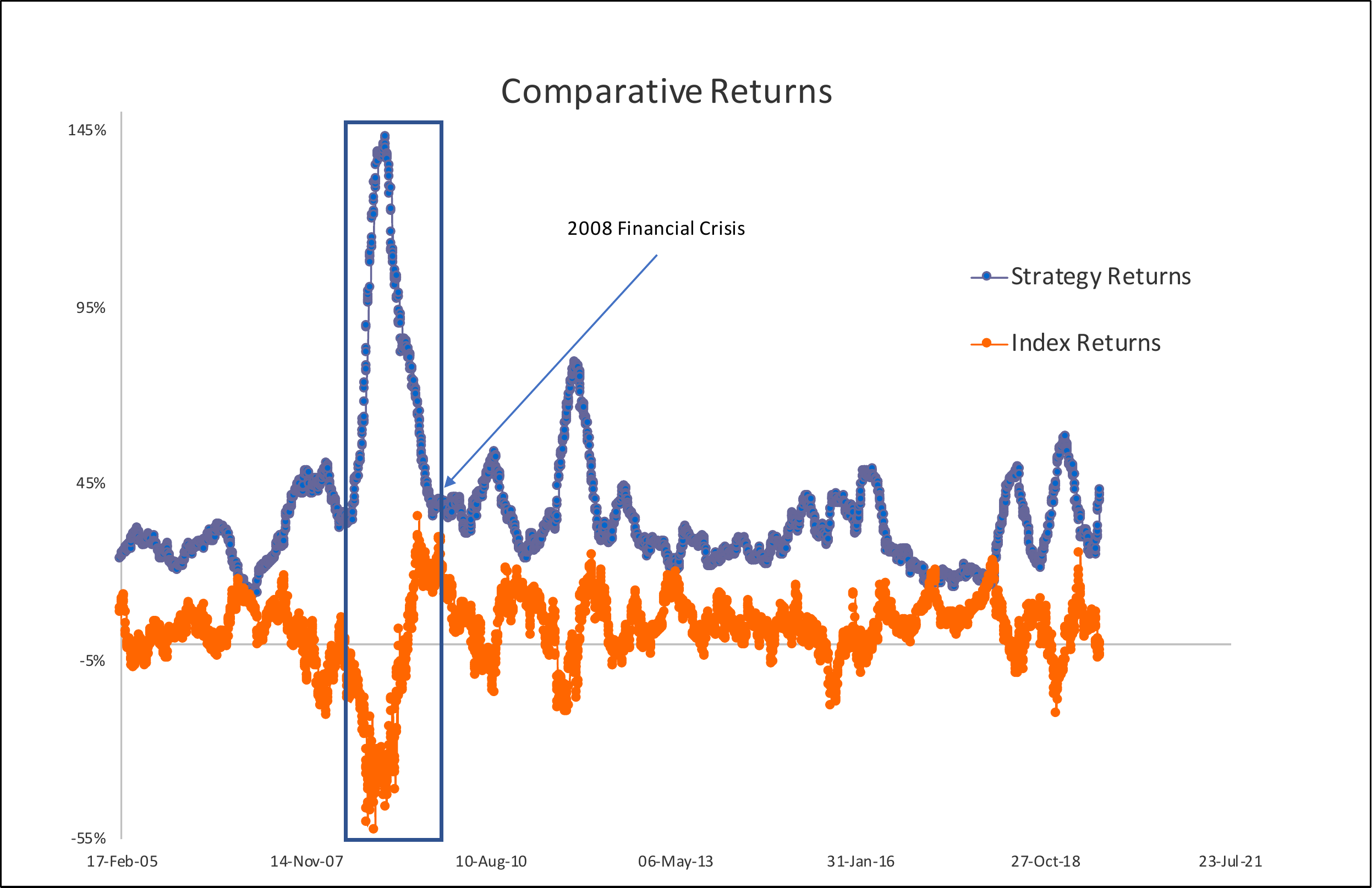}
	\centering
	\caption{Comparative returns relative to market}
\end{figure}

\section{Summary and future work}
\paragraph{}
In this article we have discussed a method that addresses some of the shortcomings of using Kelly's criterion and Sharpe ratios. This was done by using the Sortino ratio as a measure to balance risk and return. The proposed allocation and the resulting trading strategies was then simulated versus historical data from the Dow Jones Industrial Average that range from 1985 to 2019. The data set was chosen to span a long period to test how the proposed strategy weathers different economic and financial climates. The results are highly encouraging. Further work that:
\begin{enumerate}
	\item takes into account transaction costs
	\item calculates maximum drawdown and other trading risk measures
	\item estimates the potential for scalability in this approach
\end{enumerate}
\newpage
\section{Appendix}
\paragraph{}
The expression in equation (13) admits the more "elegant" form (14) which we derive here. The cumulative binomial distribution has a well known closed form in terms of the incomplete beta function \cite{CumulativeBinomialDistribution}:
\begin{equation}
\begin{split}
	\sum_{x=0}^{\alpha} \dfrac{T!}{x!(T-x)!}p^{x}(1-p)^{T-x} &=
	1 - \dfrac{T!}{\alpha!(T-\alpha-1)!}\int_{0}^{p}t^{\alpha}(1-t)^{T-\alpha}\;dt\\
	&= 1 - \dfrac{T!}{a!(-\alpha+T-1)!}B_{p}(\alpha+1, T-\alpha)
\end{split}
\tag{A.1}
\end{equation}
What we are after is : 
\begin{equation}
	\sum_{x=0}^{\alpha}(Ax + B^{2})\dfrac{T!}{x!(T-x)!}p^{x}(1-p)^{T-x} \tag{A.2}
	\label{eq:2.2}
\end{equation}

We will now show how to rewrite (A.2) in terms of (A.1). Let $\Omega$ be the differential operator: $\Omega = a\partial^{2}_{p} + b\partial_{p} + c$
We would like to find a,b and c such that:

\begin{equation}
	(Ax + B^{2})\dfrac{T!}{x!(T-x)!}p^{x}(1-p)^{T-x}=\Omega\left(\dfrac{T!}{x!(T-x)!}p^{x}(1-p)^{T-x}\right) \tag{A.3.1}
\end{equation}
where :
\begin{equation}
	A = \dfrac{1}{2}\log{(1-\Theta^{2})} \tag{A.3.2}
\end{equation}
and
\begin{equation}
	B = \dfrac{1}{2T}\log{\left(\dfrac{1+\Theta}{1-\Theta}\right)} \tag{A.3.3}
\end{equation}
By expanding the terms in equation (A.3) and applying the differential operator $\Omega$, we get the following solution for the coefficients a,b and c:
\\
\begin{equation}
a = B^{2}(p-1)^{2}p^{2} \tag{A.4.1}
\end{equation}
\begin{equation}
b = -B (p - 1) p (2 A + B (2 p (T - 1) + 1))  \tag{A.4.2}
\end{equation}
\begin{equation}
c = A^2 + 2 A B p T + (B^2) p T (p (T - 1) + 1) \tag{A.4.3}
\end{equation}
If we set :
\begin{equation}
	\alpha = T_{max} = \min{\left(T, \; 
		\dfrac{T(2\mu-\log{(1-\Theta^{2})})}{\log{\left(\dfrac{1+\Theta}{1-\Theta}\right)}}\right)} \tag{A.5}
\end{equation}
\\
and we use equations (A.1) and (A.2) we find that :

\begin{equation}
\begin{split}
\sum _{x=0}^{\alpha }\left(A+Bx \right)^2\frac{T!}{x!(T-x)!}p^x(1-p)^{T-x}
&=\Omega \left( \sum _{ x=0}^{\alpha }\frac{T!}{x!(T-x)!}p^x(1-p)^{T-x}\right)\\
&=(T-\alpha )\frac{T!}{\alpha !(T-\alpha )!}\Omega \left(\int
_0^{1-p} t^{T-\alpha -1}(1-t)^{\alpha } \textit{dt} \right) \\
&= \frac{T!}{\alpha !(T-\alpha -1)!}\left(p^{\alpha -1} (1-p)^{-\alpha +T-2} (-a
\alpha +a p (T-1)+b (p-1) p)\right.\\
&\left.\textit{            }+c B_{1-p}(T-\alpha ,\alpha +1)\right)
\end{split}
\tag{A.6}
\end{equation}

where \(B_{1-p}\) is the incomplete beta function given by :  \(B_z(a,b)=\int _0^zt^{a-1}(1-t)^{b-1}dt\). 

\newpage

\bibliography{kelly_criterion}
\end{document}